\documentclass[12pt,preprint]{aastex}

\lefthead{EVANS ET AL.} 
\righthead{HCN IN QSO Hosts}

\received{2006 July 1}
\begin{document}

\title{Dense Molecular Gas and the Role of Star Formation in the Host Galaxies of Quasi-Stellar
Objects}

\author{A. S. Evans\altaffilmark{1},
P. M. Solomon\altaffilmark{1},
L. J. Tacconi\altaffilmark{2},
T. Vavilkin\altaffilmark{1},
\& D. Downes\altaffilmark{3}}

\altaffiltext{1}{Department of Physics \& Astronomy, Stony Brook
University, Stony Brook, NY, 11794-3800: aevans@mail.astro.sunysb.edu;
psolomon@sbastk.ess.sunysb.edu; tvavilk@vulcan.ess.sunysb.edu}

\altaffiltext{2}{Max-Planck Institut fur extraterrestrische Physik,
Giessenbachstrasse, Garching D-85748, Germany: linda@mpe.mpg.de}

\altaffiltext{3}{Institut de Radio Astronomie Millimetrique, Domaine
Universitaire, F-38406 St. Martin d'Heres, France: downes@iram.fr}

\begin{abstract}

New millimeter-wave CO and HCN observations of the host galaxies of
infrared-excess Palomar Green quasi-stellar objects (PG QSOs) previously
detected in CO are presented. These observations are designed to assess the
validity of using the infrared luminosity to estimate star formation rates
of luminous AGN by determining the relative significance  of dust-heating
by young, massive stars and active galactic nuclei (AGN) in QSO hosts and
{\it IRAS} galaxies with warm, AGN-like infrared colors.  The analysis of
these data is based, in part, on evidence that HCN traces high density
($>10^4$ cm$^{-3}$) molecular gas, and that the starburst-to-HCN luminosity
ratio, $L_{\rm starburst} / L'_{\rm HCN}$, of {\it IRAS}-detected galaxies
is constant.  The new CO data provide a confirmation of prior  claims that
PG QSO hosts have high infrared-to-CO luminosity ratios, $L_{\rm IR} /
L'_{\rm CO}$, relative to {\it IRAS} galaxies of comparable $L_{\rm IR}$.
Such high $L_{\rm IR} / L'_{\rm CO}$ ratios may be due to significant
heating of dust by the QSO, or to an increased star formation efficiency in
QSO hosts relative to the bulk of the luminous {\it IRAS} galaxy
population. The HCN data show a similar trend, with the PG QSO host IZw1
and most of the warm {\it IRAS} galaxies having high $L_{\rm IR} / L'_{\rm
HCN} (>1600$) relative to the cool {\it IRAS} galaxy population for which
the median $ \left< {L_{\rm IR} / L'_{\rm HCN}} \right> _{\rm cool} \sim
890^{+440}_{-470}$.  If the assumption is made that the infrared emission
from cool {\it IRAS} galaxies is reprocessed light from embedded
star-forming regions, then high values of $L_{\rm IR}/L'_{\rm HCN}$ are
likely the result of dust heating by the AGN. Further, if the median ratio
of $L'_{\rm HCN} / L'_{\rm CO} \sim 0.06$ observed for Seyfert galaxies and
IZw1 is applied to the PG QSOs not detected in HCN, then the derived
$L_{\rm IR} / L'_{\rm HCN}$ correspond to a stellar contribution to the
production of $L_{\rm IR}$ of $\sim 7-39$\%, and star formation rates
$\sim2-37$ M$_\odot$ yr$^{-1}$ are derived for the QSO hosts.  The
corresponding values for the warm galaxies are $\sim10-100$\% and $\sim
3-220$ M$_\odot$ yr$^{-1}$. Alternatively, if the far-infrared is adopted
as the star formation component of the total infrared in cool galaxies,
i.e., $\left< L_{\rm FIR} / L'_{\rm HCN} \right> _{\rm cool} \sim L_{\rm
starburst} / L'_{\rm HCN}$, the stellar contributions in QSO hosts and warm
galaxies to their $L_{\rm FIR}$ are up to 35\% and 10\% higher,
respectively, than the percentages derived for $L_{\rm IR}$. This raises
the possibility that the $L_{\rm FIR}$ in several of the PG QSO hosts,
including IZw1, could be due entirely to dust heated by young, massive
stars. Finally, there is no evidence that the global HCN emission is
enhanced relative to CO in galaxies hosting luminous AGN.

\end{abstract}


\section{Introduction}

Over the last two decades, considerable progress has been made in
investigating the possible evolutionary connection between ultraluminous
infrared galaxy mergers (ULIGs: i.e., $L_{\rm IR} [8-1000{\rm \mu m}] \geq
10^{12}$ L$_\odot$) and quasi-stellar objects (QSOs) first proposed by
Sanders et al.  (1988a,b). These studies have transformed our view of both
the environments in which QSOs are triggered and fed, and the connection
between star formation and active galactic nuclei (AGN) activity.

Approximately 70\% of the optically-selected Palomar-Green (PG) QSO sample
have {\it IRAS}\footnote{I.e., the Infrared Astronomical Satellite.}
detections consistent with thermal emission from dust (Sanders et al.
1989). In addition, approximately a dozen low-redshift QSOs have been
detected in CO($1\to0$) to date (Sanders et al. 1988c; Barvainis et al.
1989; Alloin et al. 1992; Evans et al. 2001; Scoville et al. 2003),
indicating substantial quantities of molecular gas out of which to fuel
star formation and possibly AGN activity.

The detection of CO($1\to0$) emission does not, in itself, imply that
active star formation is occuring. Further, in galaxies known to harbor
AGN, it is unclear what fraction of the infrared emission is reprocessed
star light or AGN light, and thus using $L_{\rm IR}$ to estimate star
formation rates may not be justified. The present paper addresses these
issues by reporting the results of an HCN($1\to0$) survey of QSO
previously detected in CO. HCN has a higher dipole moment and larger
Einstein A coefficient than CO, and thus traces much denser gas ($\gtrsim
10^4$ cm$^{-3}$). In the Galaxy, HCN emission is associated with molecular
cloud cores where active star formation is occuring, and it has been
detected in a large number of infrared luminous galaxies (Gao \& Solomon
2004a). Gao \& Solomon (2004b) have shown that, for spiral and infrared
luminous galaxies spanning several orders in $L_{\rm IR}$,  the
infrared-to-HCN luminosity ratio, $L_{\rm IR} / L'_{\rm HCN}$, is
relatively constant and has smaller deviations than $L_{\rm IR}/L'_{\rm
CO}$ measured over the same $L_{\rm IR}$ range.  Gao \& Solomon conclude
that HCN is a much better tracer of gas that ultimately gives rise to the
young stellar population in these galaxies, and that the constancy of
$L_{\rm IR} / L'_{\rm HCN}$ is also proof that the $L_{\rm IR}$ is due
entirely to dust-heating by young, massive stars. With this as a working
assumption, the data presented here will be used to make a comparison
between the $L_{\rm IR} / L'_{\rm HCN}$ of the QSO host sample and
the {\it IRAS} galaxy sample. Similarities between the $L_{\rm IR} /
L'_{\rm HCN}$ ratios of the bulk of the {\it IRAS} galaxy population and that
of QSO  and {\it IRAS} galaxies with warm, AGN-like infrared colors are
expected if the $L_{\rm IR}$ of QSO hosts and warm galaxies is due to
dust-heating by young, massive stars.

This paper is divided into 6 sections. Section 2 is a summary of sample
selection.  The observing procedure and data reduction are presented in \S
3, followed by the results in \S 4.  In \S 5, the infrared, CO, and HCN
data of the PG QSO hosts are compared with that of both cool and warm {\it IRAS}
galaxies to assess the relative contribution of star formation to the
infrared emission of luminous AGN hosts.  Section 6 is a summary of the paper.
Parts of this discussion have been presented elsewhere (Evans 2003, 2005a,
2005b).  Throughout this paper, we adopt an $H_0 = 75$ km s$^{-1}$
Mpc$^{-1}$ and $q_0 = 0.5$ cosmology for consistency with Evans et al.
(2001, 2005c).

\section{Sample}

The sample was selected from Palomar-Green QSO hosts (Schmidt \& Green 1983)
detected in CO(1$\to$0) to date (see Table 1).  The sample is a subset of
the QSO hosts imaged at optical and near-infrared wavelengths by Surace, Sanders, \& Evans
(2001) and defined to have infrared-excesses (infrared-to-``big blue bump''
luminosity ratio, $L_{\rm IR} [8-1000\micron] / L_{\rm bbb}
[0.1-1.0\micron]> 0.36$) at least as high as those of ULIGs with warm,
AGN-like infrared colors (i.e., $f_{\rm 25\mu m} / f_{\rm 60\mu m} \geq
0.2$: see Sanders et al. 1988b for a discussion of the warm ULIG sample).
Within the framework of the ULIG-QSO evolutionary model, infrared-excess PG
QSO hosts represent the stage following the warm, AGN-like ULIG phase, and thus
their hosts should still show signatures consistent with
ultraluminous, gas-rich galaxy mergers (tidal tails, young star clusters,
molecular gas, etc.). Given this, they are also the most likely QSO hosts to have
the highest contribution of starlight to the production of their infrared
luminosity.  The reader is referred to Surace et al. (2001) for a more
detailed discussion of the sample selection. 

For the purposes of the study
presented here, the QSOs listed in Table 1 have three primary advantages.
First, the CO redshifts and line widths are known, thus making it
straight-forward to tune to the frequency of the HCN line and, in the
absense of a strong HCN line, to measure upper limits. Second, {\it IRAS}
and {\it ISO}\footnote{I.e., the Infrared Space Observatory.} data have
been published for these QSOs, making determinations of the infrared
luminosity possible (Table 1). Third, because of the manner in which they
are selected, they are directly comparable to the warm {\it IRAS} galaxy
sample.

\section{Observations and Data Reduction}

Observations of the redshifted HCN emission were done at the IRAM 30m
telescope during five observing periods between 2003 May and 2003 December
(see Table 2).  The 3 and 2 millimeter receivers were used in combination
with the filterbanks and an autocorrelator to provide bandwidths of 512 and
600 MHz, respectively. All QSO hosts with HCN($1\to0$) emission redshifted to
frequency accessible with the 3mm receivers were observed in HCN($1\to0$),
otherwise, HCN($2\to1$) observations were obtained with the 2mm receiver.
During observations, the pointing was monitored by observing standard
continuum sources.

Three additional sets of observations were obtained. First, all of the QSOs
were observed in CO($1\to0$) during the 2003 December and 2006 May
observing periods to verify CO redshifts and line-widths.  Second,
observations of the HCN($1\to0$) emission from Arp 220 and the CO($1\to0$)
emission from VIIZw031 were made after every other QSO observation as a
reality check of the 30m Telescope system performance. Finally, Arp220 was
observed in HCN($1\to0$) and HCN($2\to1$) to measure the ratio of two lines
(see \S 4).

The data were reduced using the IRAM data reduction package CLASS. Scans
were averaged together, and a linear baseline was subtracted from each
spectrum outside of the velocity range of the CO and HCN line emission, or,
in the cases where no HCN line was detected, outside the velocity range
where the HCN line was expected to be.  The amplitudes of each spectrum
were converted to main beam brightness temperatures, $T_{\rm mb}$, by
multiplying the spectrum by the ratio of the forward-to-beam efficiencies
of the telescope at  the specific frequency of the observation, then each
spectrum was smoothed to $\sim$ 20 km s$^{-1}$. Finally, line fluxes were
measured by numerically integrating over the channels in the line profile,
and the line widths were measured as full width at 50\% of the peak flux.
The resultant spectra are plotted in Figures 1--3.

\section{Results}

\subsection{CO}

Table 3 contains a summary of the CO emission line properties of the
sample of PG QSO hosts. The CO luminosities are calculated via,

$$L'_{\rm CO} = 2.4\times10^3 \left( S_{\rm CO} \Delta v
\over {\rm Jy~km~s}^{-1} \right) \left( D_{\rm L} \over {\rm Mpc}
\right) ^2 (1 + z)^{-1}$$
$$[{\rm K~km~s}^{-1} {\rm~pc}^2], \eqno(1)$$

\noindent
where $S_{\rm CO} \Delta v$ is the CO flux obtained by multiplying the line
intensities in Table 3 by the conversion factor for an unresolved source
(i.e., $S / T_{\rm mb} = 4.95$ J K$^{-1}$), and $D_{\rm L}$ is the
luminosity distance. The molecular gas masses, $M_{\rm H_2}$, are
calculated by adopting the conversion factor $\alpha =$ M (H$_2$)/$L'_{\rm
CO}$ = 4 M$_\odot$ (K km s$^{-1}$ pc$^2$)$^{-1}$ (see Evans et al. 2001 for a
detailed discussion of $\alpha$).

There is considerable overlap between the QSOs listed in Table 3 and recent
interferometric surveys with the Owens Valley Millimeter Array (OVRO).
Specifically, PG 0838+770, PG 1119+120, PG 1351+640, PG 1415+451, PG
1440+356, and PG 1613+658 were observed by Evans et al. (2001), and PG
2130+099 was observed as part of a volume-limited ($z<0.1$) QSO CO survey
by Scoville et al. (2003). In addition, previous single-dish measurements
of CO emission from IZw1, Mrk 1014, PG 0838+770, and PG 1613+658 exist
(Sanders et al. 1988c; Barvainis et al. 1989; Alloin et al. 1992).

The general shape and width of the CO lines are consistent with previously
published CO spectra of these QSOs.  For the QSOs with broad ($> 200$ km
s$^{-1}$) CO emission lines, the CO velocity width measured at half the
maximum intensity, $\Delta v_{\rm FWHM}$, agree to within 20\% of prior
measurements. The agreement degrades to 40\% for the narrow ($<100$ km
s$^{-1}$) CO line QSOs, the larger discrepancy resulting from the
difficulty in accurately measuring $\Delta v_{\rm FWHM}$ for narrow
emission lines after rebinning the data. From these new measurements, an
average CO emission line of $\Delta v_{\rm FWHM} \sim 280\pm150$ km
s$^{-1}$ is estimated, as compared to the value of $300\pm90$ km s$^{-1}$
measured for ULIGs (see Solomon et al. 1997 and Evans et al. 2005c).

With the exception of PG 1613+658, the new measurements of main beam CO
line fluxes, $S_{\rm CO} \Delta v$, are on average 40\% lower than the OVRO
data presented in Evans et al. (2001).  The new CO line flux measurement of
PG 1613+658, which is the PG QSOs with the highest signal-to-noise CO
detection in Evans et al. (2001), agrees to within 4\% with the OVRO data.
Finally, the CO line flux of PG 2130+099 agrees to within 10\% of the
measurement made by Scoville et al.  (2003).

While previous single-dish $S_{\rm CO} \Delta v$ measurements of IZw1 and
PG 1613+658 are consistent to within 20\% of the measurements presented
here, single-dish measurements of PG 0838+770 and Mrk 1014 are up to 300\%
higher than the present measurements. The discrepancies are undoubtedly a
result of calibration errors and low signal-to-noise data.

\subsection{HCN}

Table 3 also summarizes the HCN emission line properties of the QSO sample
and Arp 220.  Only the PG QSO host IZw1 was detected in HCN($1\to0$); the
shape and width of the HCN line are consistent with the CO($1\to0$)
profile. For the remaining PG QSO hosts, the HCN line intensities were
measured by numerically integrating over the channels expected to contain
HCN emission based on the CO($1\to0$) observations. In addition, 3$\sigma$
upper limits to the HCN line intensities were derived via,

$${T_{\rm HCN} \Delta v} < {{3 T_{\rm rms} \Delta v_{\rm FWZI}} \over
{\sqrt {\Delta v_{\rm FWZI} / \Delta v_{\rm res}}}}\
~~~[{\rm K~km~s}^{-1}], \eqno(2)$$

\noindent
where $\Delta v_{\rm FWZI}$ is the full width at zero intensity velocity of
the CO emission line of the QSO host, and $T_{\rm rms}$ is the root-mean-squared
main beam temperature of the HCN spectral data for a velocity resolution of
$\Delta v_{\rm res}$. Line luminosities were calculated via,

$$L'_{\rm HCN(J \to J-1)} = 4.145\times10^3 \left( S_{\rm HCN} \Delta v
\over {\rm Jy~km~s}^{-1} \right) \left( D_{\rm L} \over {\rm Mpc}
\right) ^2 J^{-2} (1 + z)^{-1}$$
$$[{\rm K~km~s}^{-1} {\rm~pc}^2]. \eqno(3)$$
 
The HCN(1$\to$0) lines of two of the PG QSOs (PG 0838+770 and PG 1613+658)
are redshifted to frequencies below the range accessible with the 3mm
receivers. In both cases, the 2mm receivers were tuned to the frequencies
corresponding to the redshifted HCN(2$\to$1) emission line. Both
observations resulted in upper limits, which were converted to HCN(1$\to$0)
upper limits by adopting the $L'_{\rm HCN(1\to0)} / L'_{\rm HCN(2\to1)}$ (=2.997)
line ratio measured for Arp 220 (Figure 3).

\section{Discussion}

Approximately one quarter of the PG QSOs at $z <  0.3$ have been detected
in CO(1$\to$0), and $\sim 50-70$\% show evidence of spiral disks,
merger-like tidal features, and/or luminous star clusters (e.g., Stockton
\& MacKenty 1983; Surace et al. 2001 and references therein; Veilleux et
al. 2006; Guyon et al. 2006). The presence of star-forming molecular gas and luminous
star clusters, combined with infrared luminosities in excess of $10^{11}$ L$_\odot$, 
are compelling indications that star formation is an important component of 
activity in these galaxies, and it raises the possibility that
a significant fraction of their thermal infrared emission may be reprocessed
stellar light.

Given such evidence, an obvious question to ask is - what are the star
formation rates of the PG QSO hosts?  As a naive first attempt, the assumption
will be made that the infrared luminosity of the host is due entirely to
dust-heating by the starburst population, and that $\tau >> 1$ towards the
star-forming regions.  Doing so yields star formation rates of

$${SFR \sim 1.76 \times10^{-10} \left( {L_{\rm IR} \over {\rm L_\odot} }\right)
\left( {\delta \over 0.75} \right)
[{\rm M_\odot ~ yr^{-1}}] \sim 20 - 420~{\rm M_\odot ~ yr^{-1}}}, \eqno(4)$$

\noindent 
with an average star formation rate of $\left< SFR \right> \sim 90\pm 125$
M$_\odot$ yr$^{-1}$ and the median of 40 M$_\odot$ yr$^{-1}$ (Table 1).
Equation 4 is a modified version of Equation 4 in Kennicutt (1998), where
the factor $\delta$ ($= 0.75\pm0.07$) is the average far
infrared-to-infrared luminosity ratio, $L_{\rm FIR} (40-500{\rm \mu m}) /
L_{\rm IR} (8-1000{\rm \mu m})$, calculated for the {\it IRAS} galaxies in
the Revised Bright Galaxy Sample ($\sim 600$ galaxies: Sanders et al.
2003).  These star formation rates are likely upper limits -- it is
difficult to imagine a geometry in which light from the AGN does not
intersect (and thus heat) a significant fraction of dust in the host
galaxy. Indeed, models have been developed that can account for all of
infrared emission in QSO hosts by an AGN surrounded by a warped or clumpy torus
(Sanders et al. 1989; see review in Haas et al. 2003; Elitzur et al. 2004).

\subsection{$L_{\rm IR} / L'_{\rm CO}$: Star Formation Efficiencies versus 
Dust Heating by the AGN}

To examine the importance of dust-heating by young, massive stars to the
production of $L_{\rm IR}$ in their hosts, consider first the $L'_{\rm CO}$ and
$L_{\rm IR}$ of QSO hosts relative to that of {\it IRAS}-detected galaxies
(Figure 4: see also Evans et al. 2001, 2005c).  The new $L'_{\rm CO}$
measurements of the infrared-excess QSO hosts (Table 3) are plotted in Figure
4\footnote{Note that Evans et al. (2001) use $L_{\rm IR} (1 - 1000{\rm \mu
m})$ for QSO hosts instead of the commonly used $L_{\rm IR} (8 - 1000{\rm
\mu m})$. The $L_{\rm IR} (1 - 1000{\rm \mu m})$ in Evans et al. (2001)
were calculated by numerically integrating the spectral energy distributions
in Sanders et al. (1989). Here, we use $L_{\rm IR} (8 - 1000{\rm \mu m})$,
which leaves out the significant contribution of hot dust in the 1 -
8$\mu$m range, but provides better consistency with the other datasets
plotted.} -- additional QSO data are compiled from Scoville et al. (2003).
In order to provide a more meaningful comparison with the QSO hosts, the {\it
IRAS} galaxies have been separated into those with warm, Seyfert-like
(i.e., $f_{\rm 25\mu m} / f_{\rm 60\mu m} \geq 0.20$) and cool ($f_{\rm
25\mu m} / f_{\rm 60\mu m} < 0.20$) infrared colors.  There is significant
evidence that most warm {\it IRAS} galaxies host AGN, but there is not clear
evidence that cool {\it IRAS} galaxies, as a class, have luminous AGN.
Given this, the cool galaxies will be treated as galaxies powered by
starbursts.

The new CO data show similar results to what is shown in Evans et al.
(2001, 2005c) -- the QSO hosts have low $L'_{\rm CO}$ for their $L_{\rm IR}$
relative to the bulk of cool {\it IRAS}  galaxies. In other words, QSO hosts
have high $L_{\rm IR} / L'_{\rm CO}$ relative to cool {\it IRAS} galaxies
with comparable $L_{\rm IR}$. The warm galaxies with $L_{\rm IR} >
10^{10.0}$ L$_\odot$ show a similar trend in $L_{\rm IR} / L'_{\rm CO}$
relative to cool {\it IRAS} sample, though not as dramatically as the QSO host
population.

To interpret the data plotted in Figure 4, consider 

$${L_{\rm IR} \over L'_{\rm CO}} \sim {L_{\rm starburst} + \epsilon L_{\rm
AGN} \over ( M_{\rm H_2} / \alpha )},\eqno(5)$$

\noindent
where $L_{\rm starburst}$ and $L_{\rm AGN}$ are the luminosities of the
starburst and AGN, respectively, and $\epsilon$ is the fraction of the AGN
light absorbed by dust and reradiated in the thermal infrared. In the limit
where either $L_{\rm AGN} = 0$ or $\epsilon = 0$, Equation 5 is simply the
star formation efficiency. I.e., if $\tau >>1$ towards the starburst
population which is producing stars in steady state, $L_{\rm IR} / L'_{\rm
CO}$ is a measure of total energy output of the starburst population per
unit of fuel available to form new stars. If $\epsilon = 0$ for QSO hosts, then
the high $L_{\rm IR} / L'_{\rm CO}$ is an indication that they are
producing stars extremely efficiently from the available molecular gas.
Indeed, one would be forced to conclude that the star formation efficiency
is preferentially higher in galaxies hosting luminous AGN. The alternative
interpretation is that $\epsilon > 0$, and thus that $\epsilon L_{\rm AGN}$
contributes significantly to $L_{\rm IR}$.

\subsection{Dense Molecular Gas and Star Formation Rates}

The conclusions stated in the above discussion are only valid if CO is a
robust tracer of star formation in these galaxies, i.e., if the CO emission
is tracing most, if not all, of the molecular hydrogen
that is actively forming stars.  Given that HCN appears to be more closely
coupled with star formation in starburst galaxies than CO (see \S 1),
applying a similar analysis to HCN as done in \S 5.1 for CO will likely
yield a more physically meaningful result. 

Figure 5 is a plot of $L_{\rm
IR} / L'_{\rm HCN}$ versus $L_{\rm IR}$ of the QSO hosts, {\it IRAS}
galaxies, and two high-redshift AGN recently detected in HCN($1\to0$).
Again, the {\it IRAS} galaxies have been separated into those with warm and
cool infrared colors. If the dust in the cool {\it IRAS} galaxies are
heated primarily by the starburst population ($\epsilon \sim 0$), then the
median $L_{\rm IR} / L'_{\rm HCN}$ of the cool {\it IRAS} sample will be
adopted as the ratio at which $L_{\rm IR} = L_{\rm starburst}$. I.e.,

$$\left<{{L_{\rm IR} \over L'_{\rm HCN}}}\right> _{\rm cool} \sim 
{L_{\rm starburst} \over ( M_{\rm H_2} / \alpha _{\rm HCN} )} 
\sim 890^{+440}_{-470} {\rm ~L_\odot~(K ~km ~s^{-1}~ pc^2)^{-1}},\eqno(6)$$

\noindent
where $\alpha _{\rm HCN}$ is the $L'_{\rm
HCN}$-to-M$_{H_2}$ mass conversion factor and the limits on 
$\left<{{L_{\rm IR} / L'_{\rm HCN}}}\right> _{\rm cool}$ 
represent the range in which 67\% (i.e., $1\sigma$) of the data points 
around the median are contained.

The upper portion of Figure 5 (i.e., high $L_{\rm IR} / L'_{\rm HCN}$) is
mostly populated with the warm galaxies and AGN hosts.  Specifically, the
QSO host IZw1, the two high redshift AGN hosts, and two of the eight warm
{\it IRAS} galaxies have much higher $L_{\rm IR} / L'_{\rm HCN}$ ($>2200$)
than the cool galaxy population (i.e., low values of $L'_{\rm HCN}$ for
their $L_{\rm IR}$).  These galaxies clearly have a significant
contribution to their $L_{\rm IR}$ from AGN activity. Further, the fact
that both of the cool {\it IRAS} galaxies within this $L_{\rm IR}/L'_{\rm
HCN}$ range (Arp 299 and Mrk 273, for which $L_{\rm IR} / L'_{\rm HCN} \sim 3000$) 
have strong X-ray signatures of embedded
AGN (Della Ceca et al.  2002; Xia et al. 2002; Zesas, Ward, \& Murray 2003)
also supports the idea that high $L_{\rm IR} / L'_{\rm HCN}$ are the result of
significant dust-heating by AGN.  Four of the eight warm galaxies have $L_{\rm IR} /
L'_{\rm HCN}$ in the range 1600--2000, and thus overlap the high end of the
bulk of the cool galaxy distribution. These galaxies, which include the
well-studied ULIG Mrk 231, might have an equal mixture of  AGN and
starburst emission as viewed in the thermal infrared, however, they could
also be almost entirely powered by star formation as their $L_{\rm IR} /
L'_{\rm HCN}$ is not outside the range of that  found in cool galaxies.
Finally, two of the eight warm galaxies have $L_{\rm IR} / L'_{\rm HCN}$
very near the median of the cool galaxy population, illustrating the fact that
some fraction of the warm galaxy population have ``normal'' $L_{\rm IR} / L'_{\rm HCN}$. 
Whether or not the PG QSO hosts not detected in HCN 
have high $L_{\rm IR} / L'_{\rm HCN}$ similar to IZw1
cannot be addressed with the data in Figure 5.

While only one of the eight PG QSO hosts was detected in HCN, and the
$L_{\rm IR} / L'_{\rm HCN}$ lower limits of the undetected QSO hosts are
not sufficiently constraining to be conclusive, the $L_{\rm IR}/L'_{\rm
HCN}$ ratios of the undetected hosts  can possibly be inferred by making
use of the larger number of HCN-detected Seyferts in the {\it IRAS} galaxy
sample. Figure 6 is a plot of $L'_{\rm HCN} / L'_{\rm CO}$ versus $L_{\rm
IR}$, and is a modified version of Figure 4 in Gao \& Solomon (2004b).
Here, the {\it IRAS} galaxies have been plotted in terms of their optical
emission-line classifications (i.e., Seyferts, LINERs\footnote{Low
Ionization Nuclear Emission Region galaxies}, and HII region-like
galaxies).  In this Figure, there is an increased spread in $L'_{\rm HCN} /
L'_{\rm CO}$ values for galaxies with $L_{\rm IR} > 10^{11}$ L$_\odot$,
which Gao \& Solomon (2004b) attribute to the increase in the fraction of
dense gas in luminous infrared galaxies corresponding to the increase in
starburst activity.  Eleven of the 13 (85\%) Seyferts in Figure 6 have
$L'_{\rm HCN} / L'_{\rm CO} <0.1$.  In addition, IZw1 has a $L'_{\rm HCN} /
L'_{\rm CO}$ ratio within the distribution of Seyfert data points at
$L'_{\rm HCN} / L'_{\rm CO} <0.1$. Given this, it is a reasonable
assumption that the PG QSO hosts in the present sample have low $L'_{\rm
HCN} / L'_{\rm CO}$ similar to the bulk of the Seyfert galaxies in Figure
6. Thus, for the PG QSO hosts, the median $L'_{\rm HCN} / L'_{\rm CO}$ of
Seyferts and IZw1 in Figure 6, i.e.

$$\left< {L'_{\rm HCN} \over L'_{\rm CO}} \right> _{\rm Seyfert} \sim 0.06^{+0.03}_{-0.04}, \eqno(7)$$

\noindent
is adopted, where the uncertainties span the range of $L'_{\rm HCN} /
L'_{\rm CO}$ for the 11 low $L_{\rm HCN} / L'_{\rm CO}$ Seyfert galaxies.
A direct estimation of $L_{\rm IR} / L'_{\rm HCN}$ for the QSO hosts can
then be made from the $L_{\rm IR} / L'_{\rm CO}$ of the QSO hosts listed in
Table 3.  The resultant $L_{\rm IR} / L'_{\rm HCN}$ ratios are listed in
Table 4, and the data are plotted in Figure 7. The stellar contributions to
the production of $L_{\rm IR}$ of PG QSOs hosts is thus

$$\Delta _{\rm starburst} ({\rm IR}) \sim 
1.76\times10^6 \left( {\left< L_{\rm IR} / L'_{\rm HCN} \right> _{\rm cool} \over 890} \right)
\left( { \left< L'_{\rm HCN} / L'_{\rm CO} \right> _{\rm Seyfert}  \over 0.06} \right)$$
 
$$\left( { {\rm L_\odot~ [K~ km~ s^{-1}~ pc^2]^{-1} \over L_{\rm IR} / L'_{\rm CO}} } \right)
~{\rm [\%]} \sim
7-39\%, \eqno(8)$$ 

\noindent
and the star formation rates are 2--37 M$_\odot$ yr$^{-1}$ (median = 10
M$_\odot$ yr$^{-1}$). Note, however, that given the uncertainties in
Equation 7, the possibility that massive, young stars produce 60\% of
$L_{\rm IR}$ cannot be ruled out for 50\% of the QSOs plotted in Figure 7.
By comparison, the corresponding values for the warm galaxies are $\sim
10-100$\% and $\sim 3-220$ M$_\odot$ yr$^{-1}$ (median = 20 M$_\odot$
yr$^{-1}$), respectively.

Clearly, a significant fraction of the cool {\it IRAS} galaxies do harbor
AGN which may or may not contribute significantly to the heating of dust in
their host galaxy. Thus, as a double-check of the results derived from
Figure 7, the Figure has been replotted with the cool {\it IRAS} galaxies
optically classified as Seyferts and LINERs\footnote{Note that, while some
of the LINERs are likely primarily starburst galaxies, some of them
have strong evidence of embedded AGN.
Thus, the conservative approach of omitting
all of the LINERs has been taken for this exercise.} omitted (Figure 8).
Galaxies with transitional spectra (e.g., galaxies classified both as
LINERs and HII-region like galaxies via different line diagnostics) have
been retained, and galaxies with no published classification have also been
omitted. Finally, Arp 299, which has an HII region-like optical classification but is
known via X-ray observations to harbor an AGN, has been omitted. 
The main outcome of plotting only the remaining HII region-like galaxies and
transitional galaxies is the removal of the outliers in the $ L_{\rm IR} /
L'_{\rm HCN} $ distribution. The value of $\left< L_{\rm IR} / L'_{\rm HCN}
\right> _{\rm cool} $ ($= 910\pm430$) remains essentially unchanged.

\subsection{Dense Molecular Gas and $L_{\rm FIR}$}

Another obvious issue to consider is whether the above results differ if
the far-infrared emission from cool galaxies is considered as the star
formation component of the total infrared emission, i.e., if $\left< L_{\rm
FIR} / L'_{\rm HCN} \right> _{\rm cool} \sim L_{\rm starburst} / L'_{\rm
HCN}$.  Figure 9 is a plot of $L_{\rm FIR} / L'_{\rm HCN}$ versus $L_{\rm
FIR}$ of the same samples plotted in Figure 7.  The use of $L_{\rm FIR}$
instead of $L_{\rm IR}$ reduces the disparity between the PG QSO and
high-$z$ AGN hosts data points relative to the {\it IRAS}  galaxy data
points, and shows the possibility that the far-infrared emission in several
PG QSO hosts could be due almost entirely to dust heated by young, massive
stars (see Table 4).  In particular, IZw1 has a $L_{\rm FIR} / L'_{\rm
HCN}$ very near the median value of the cool {\it IRAS} galaxy population.
In the case of the warm galaxies, the median $\left< L_{\rm FIR} / L'_{\rm
HCN} \right> _{\rm warm}$ ($\sim 1220^{+860}_{-810}$) is 33\% lower than
$\left< L_{\rm IR}/L'_{\rm HCN} \right> _{\rm warm}$, however, the median
value of the cool {\it IRAS} galaxies, $\left< L_{\rm FIR}/L'_{\rm HCN}
\right> _{\rm cool} \sim 660^{+370}_{-330}$, is 25\% lower than $\left<
L_{\rm IR}/L'_{\rm HCN} \right> _{\rm cool}$. Thus, the relative spacing of
the warm and cool galaxy data points are only marginally changed by using
$L_{\rm FIR}$.  As a result, the contribution of dust-heating by young,
massive stars to the production of $L_{\rm FIR}$ for the PG QSO hosts and
warm galaxies increases by 1-35\% and 1--12\%, respectively,
over the contribution derived for $L_{\rm IR}$ (see Table 4). These results
reflect the fact that the average $ L_{\rm FIR} / L_{\rm IR}$ for the cool,
warm, and PG QSO galaxies in Figure 8 are $0.77\pm0.05$, $0.64\pm0.11$, and
$0.44\pm0.10$, respectively.

\subsection{Comparison with Recent Estimates of Star Formation Rates of PG QSO
Hosts}

The general conclusion that the PG QSO host star formation rates in
Equation 4 may, in some cases, be overestimates is consistent with a recent
study of the star formation rates of QSO hosts by Ho (2005). In this study,
star formation rates were derived by first assuming that the [O II]
$\lambda 3727$ / [O III] $\lambda 5007$ line ratio in QSOs is the same as
the constant value measured for the narrow-line regions of Seyfert
galaxies, then attributing excesses in the ratio to star formation within
the host. Doing so yields [O II] $\lambda 3727$-derived star formation
rates of $\sim 0.6-20$ M$_\odot$ yr$^{-1}$ - i.e., significantly lower than
what would be derived based solely on the assumption that the $L_{\rm IR}
\sim L_{\rm starburst}$, and is consistent with the low end of the range in
star formation rates estimated for individual hosts in Table 4. Note,
however, that the [O II] $\lambda 3727$ emission line is subject to
significant extinction by dust, and thus the Ho (2005) estimates are likely
lower limits.

Another study (Schweitzer et al. 2006) addresses the issue of the nature of
far-infrared emission in PG QSO hosts by applying the 7.7 $\mu$m PAH-to-far
infrared luminosity ratio  ($L_{\rm PAH} / L_{\rm FIR}$) versus [Ne II]
12.5$\mu$m-to-far infrared luminosity ratio ($L_{\rm [Ne II]} / L_{\rm
FIR}$) diagnostic of starburst galaxies to the QSO host sample.  Starburst
galaxies are observed to have strong PAH and weak [Ne II] relative to
AGN-dominated hosts because the AGN efficiently destroys PAH and is able to
produce extended semi-ionized regions in which neon is collisionally
excited. Schweitzer et al.  conclude that 30\% or more of the far-infrared
emission to due to dust heating by young, massive stars, relative to the
median percentage of $42^{+56}_{-35}$\% estimated from the present analysis
of the HCN data.

\subsection{The Effects of the AGN on HCN Emission}

Throughout this paper, the assumption has been made that HCN is
collisionally excited by molecular hydrogen. Recent papers have challenged
this claim by stating that, in galaxies harboring AGN, X-ray emission from
the AGN enhances HCN emission (e.g., Kohno et al. 2003; Gracia-Carpio et
al. 2006).  The high $L_{\rm IR} / L'_{\rm HCN}$, and thus the low $L'_{\rm
HCN}$, of most of the PG QSO hosts, high-$z$ AGN hosts,  and warm galaxies
relative to cool {\it IRAS} galaxies (Figure 7) clearly contradicts this
hypothesis.

To examine this issue another way, consider again the plot of $L'_{\rm HCN}
/ L'_{\rm CO}$ versus $L_{\rm IR}$ of the QSO hosts and {\it IRAS} galaxies
(Figure 6).  In this Figure, IZw1 and most of the Seyfert galaxies
have $L'_{\rm HCN} / L'_{\rm CO} < 0.1$, consistent with
the distribution of the HII region-like galaxies.  Two of the Seyfert galaxies have
$L'_{\rm HCN} / L'_{\rm CO} > 0.1$, however, they are still within the
distribution of $L'_{\rm HCN} / L'_{\rm CO} < 0.1$ of HII region-like and
LINER galaxies. Thus, there is no evidence that the global HCN emission is
enhanced relative to CO by the presence of a powerful AGN.

\section{Conclusions}

New CO and HCN observations of a sample of $z<0.17$ optically-selected PG
QSO hosts with infrared excesses are presented. These data have been compared
with similar data of {\it IRAS}-detected galaxies to assess the validity of
using the infrared luminosity to estimate star formation rates in luminous
AGN hosts. The following conclusions are reached:

1. The new CO measurements confirm the results presented in Evans et al.
(2001) -- that QSO host as a class may have low values of $L'_{\rm CO}$ for
their $L_{\rm IR}$ relative to {\it IRAS} galaxies, or relatively high
values of $L_{\rm IR} / L'_{\rm CO}$. Such high $L_{\rm IR} / L'_{\rm CO}$
ratios may be evidence that AGN contribute significantly to the heating
of dust in their host galaxies, or that they have high star formation
efficiencies.

2. A comparison of the QSO hosts and {\it IRAS} galaxies surveyed in
HCN(1$\to$0) to date show the upper distribution of $L_{\rm IR} / L'_{\rm
HCN}$ ($>1600$) to be mostly populated by warm {\it IRAS} galaxies and AGN
hosts.  This distribution overlaps at the low end with cool {\it IRAS}
galaxies for which $\left< {L_{\rm IR} / L'_{\rm HCN}} \right> _{\rm cool}
\sim 890^{+440}_{-470}$ L$_\odot$ (K km s$^{-1}$ pc$^2$)$^{-1}$.  There are
also 2 (out of 8) warm galaxies with $L_{\rm IR} / L'_{\rm HCN}$ comparable
to the median $L_{\rm IR} / L'_{\rm HCN}$ of cool galaxies.  An enhanced
$L_{\rm IR} / L'_{\rm HCN}$ relative to cool {\it IRAS} galaxies thus appears to be an
indication that an AGN is contributing significantly to heating the dust.
However, the possibility that the dust in most of the QSO hosts and warm
galaxies are heated primarily by young, massive stars cannot be entirely
ruled out.  If the median ratio of $L'_{\rm HCN} / L'_{\rm CO} \sim 0.06$
observed for Seyfert galaxies and the PG QSO host IZw1 is applied to the PG
QSOs not detected in HCN, then the derived $L_{\rm IR} / L'_{\rm HCN}$
correspond to a stellar contribution to the production of $L_{\rm IR}$ of
$\sim 7-35$\% and star formation rates $\sim2-37$ M$_\odot$ yr$^{-1}$ are 
derived for the QSO hosts.  The corresponding values for the warm galaxies
are $\sim10-100$\% and $\sim 3-220$ M$_\odot$ yr$^{-1}$.

3. The average $\left< L_{\rm FIR} / L_{\rm IR} \right>$ of the cool, warm,
and PG QSO galaxies considered in this study are $0.77\pm0.05$,
$0.64\pm0.11$, and $0.44\pm0.10$, respectively.  Thus, the disparity
between the QSO (and high-redshift AGN) hosts and the cool galaxies is
diminished by comparing the $L_{\rm FIR}$ of QSO hosts and cool galaxies.
If the far-infrared is adopted as the star formation component of the total
infrared in cool galaxies, i.e., $\left< L_{\rm FIR} / L'_{\rm HCN} \right>
_{\rm cool} \sim L_{\rm starburst} / L'_{\rm HCN}$, then five of the nine
QSO hosts, including IZw1, potentially have all of their $L_{\rm FIR}$ generated by dust
heated by massive stars. The stellar contributions to $L_{\rm FIR}$ in QSO
hosts are up to 35\% higher than those derived for $L_{\rm IR}$, but the
warm galaxies have contributions only 10\% higher than those derived for
$L_{\rm IR}$.

4. The PG QSO host IZw1 and the {\it IRAS} galaxies with Seyfert optical
classifications have $L'_{\rm HCN} / L'_{\rm CO}$ distributions comparable
to HII-region like galaxies.  It is thus unlikely that the global HCN
emission in galaxies harboring luminous AGN is enhanced relative to CO by
radiation from the AGN.

A HCN survey of a more statistically complete sample of  {\it IRAS}
galaxies and PG QSO hosts is required to assess the robustness of many of
these results. The apparent faintness of HCN emission in PG QSO hosts, and
in many warm galaxies, necessitates the use of a larger millimeter-wave
telescope. Such observations may be possible with the soon-to-be
commissioned 50m-diameter Large Millimeter Telescope (LMT), or the
100m-diameter Green Bank Telescope (GBT). Frequencies as high as 90 GHz
should be accessible with the GBT in a few years.

\acknowledgements

We thank the telescope operators and staff of the IRAM 30m telescope for
their support both during and after the observations were obtained, and the
anonymous referee for critical comments that greatly improved the accuracy
and clarity of the paper. ASE also thanks J. Mazzarella and J. Mulchaey for
useful discussions and assistance.  ASE was supported by NSF grant AST
02-06262.  This research has made use of the NASA/IPAC Extragalactic
Database (NED) which is operated by the Jet Propulsion Laboratory,
California Institute of Technology, under contract with the National
Aeronautics and Space Administration.

%\vfill\eject

\centerline{Figure Captions}

%\vskip 0.3in

%\noindent
%Figure 1. 
\figcaption{IRAM 30m spectra of CO($1\to0$) emission from nine PG QSO hosts with
infrared-excesses. The intensity scale is in units of main beam brightness
temperature. A linear baseline has been subtracted from each spectrum; the
baseline subtraction is performed outside of the velocity range of the
emission lines. The zero velocity corresponds to the redshifts listed in
Table 1.}

%\noindent
%Figure 2. 
\figcaption{IRAM 30m spectra of HCN($1\to0$) and HCN($2\to1$) emission from
the QSO host sample. The intensity scale is in units of main beam
brightness temperature. A linear baseline has been subtracted from each
spectrum; the baseline subtraction is performed outside of the velocity
range where the emission line is expected to be based on the position and
velocity width of CO($1\to0$) emission from these QSO hosts.}

%\noindent
%Figure 3. 
\figcaption{IRAM 30m spectra of HCN($1\to0$) and HCN($2\to1$) emission from
the ultraluminous infrared galaxy Arp 220. The intensity scale is in units
of main beam brightness temperature. A linear baseline has been subtracted
from each spectrum; the baseline subtraction is performed outside of the
velocity range of the emission lines. The zero velocity corresponds to a
redshift of 0.0181.}

%\noindent
%Figure 4. 
\figcaption{$L'_{\rm CO}$ vs. $L_{\rm IR}$ of the PG QSO hosts and a sample of
{\it IRAS}-detected galaxies. References for the data are as follows:
infrared galaxies, Mazzarella et al. (1993a) and Solomon et al. (1997); PG
QSO hosts, this paper and Scoville et al. (2003).}

%\noindent
%Figure 5. 
\figcaption{$L_{\rm IR} / L'_{\rm HCN}$ vs. $L_{\rm IR}$ of the PG QSO hosts,
a sample of local {\it IRAS} galaxies, and two high-$z$ AGN (FSC 10214+4724
and the Cloverleaf Quasar). Filled squares with arrows represent 3$\sigma$
lower $L_{\rm IR} / L'_{\rm HCN}$ limits of the PG QSOs not detected in
HCN. The dashed line is $L_{\rm IR} / L'_{\rm HCN} = 890$ L$_\odot$ (K km
s$^{-1}$ pc$^2$)$^{-1}$ (see \S 5.1) -- the error bar associated with the
dashed line encloses 67\% ($1\sigma$) of the cool galaxy data points.  With
the exception of NGC 4418 (Imanishi et al. 2004), Mrk 273 and IR 17208-0014
(Imanishi, Nakanishi, \& Kohno 2006), and Arp 220 (this paper), the {\it
IRAS} galaxy HCN data are compiled from Gao \& Solomon (2004a,b). The
F10214+4724 and Cloverleaf data are from Vanden Bout, Solomon, \& Maddalena
(2004) and Solomon et al. (2003), respectively.}

%\noindent
%Figure 6. 
\figcaption{The $L'_{\rm HCN} / L'_{\rm CO}$ versus $L_{\rm IR}$ of IZw1 and
{\it IRAS} galaxies. The {\it IRAS} galaxy data are plotted by their
optical emission-line classification. The warm {\it IRAS} galaxies are
encircled.}

%\noindent
%Figure 7. 
\figcaption{Same as Figure 5, except the estimated $L_{\rm IR} / L'_{\rm
HCN}$ (see \S 5.2) of the PG QSO hosts not detected in HCN are plotted and
the y-axis upper limit has been increased. The
vertical bars associated with the PG QSO host data points cover the range
in $L_{\rm IR} / L'_{\rm HCN}$ that result from the uncertainties in
Equation 7.}

%\noindent
%Figure 8. 
\figcaption{Same as Figure 7, except only the cool {\it IRAS} galaxies
optically classified as HII-region like galaxies or transition galaxies are
plotted (Arp 299 has also been omitted - see text). 
The warm galaxies and PG QSO host data points remain unchanged.
Optical emission-line classifications are obtained from Veilleux et al.
(1995), Ho, Filippenko, \& Sargent (1977), Mazzarella \& Boroson (1993b),
and the NASA Extragalactic database (NED).}

%\noindent
%Figure 9. 
\figcaption{$L_{\rm FIR} / L'_{\rm HCN}$ vs. $L_{\rm FIR}$ of the PG QSO
hosts, a sample of local {\it IRAS} galaxies, and two high-$z$ AGN (FSC
10214+4724 and the Cloverleaf Quasar). The dashed line is $L_{\rm FIR} /
L'_{\rm HCN} = 660$ L$_\odot$ (K km s$^{-1}$ pc$^2$)$^{-1}$ (see \S 5.3) --
the error bar associated with the dashed line encloses 67\% ($1\sigma$) of
the cool galaxy data points.}

%\clearpage

%\newpage
\plotone{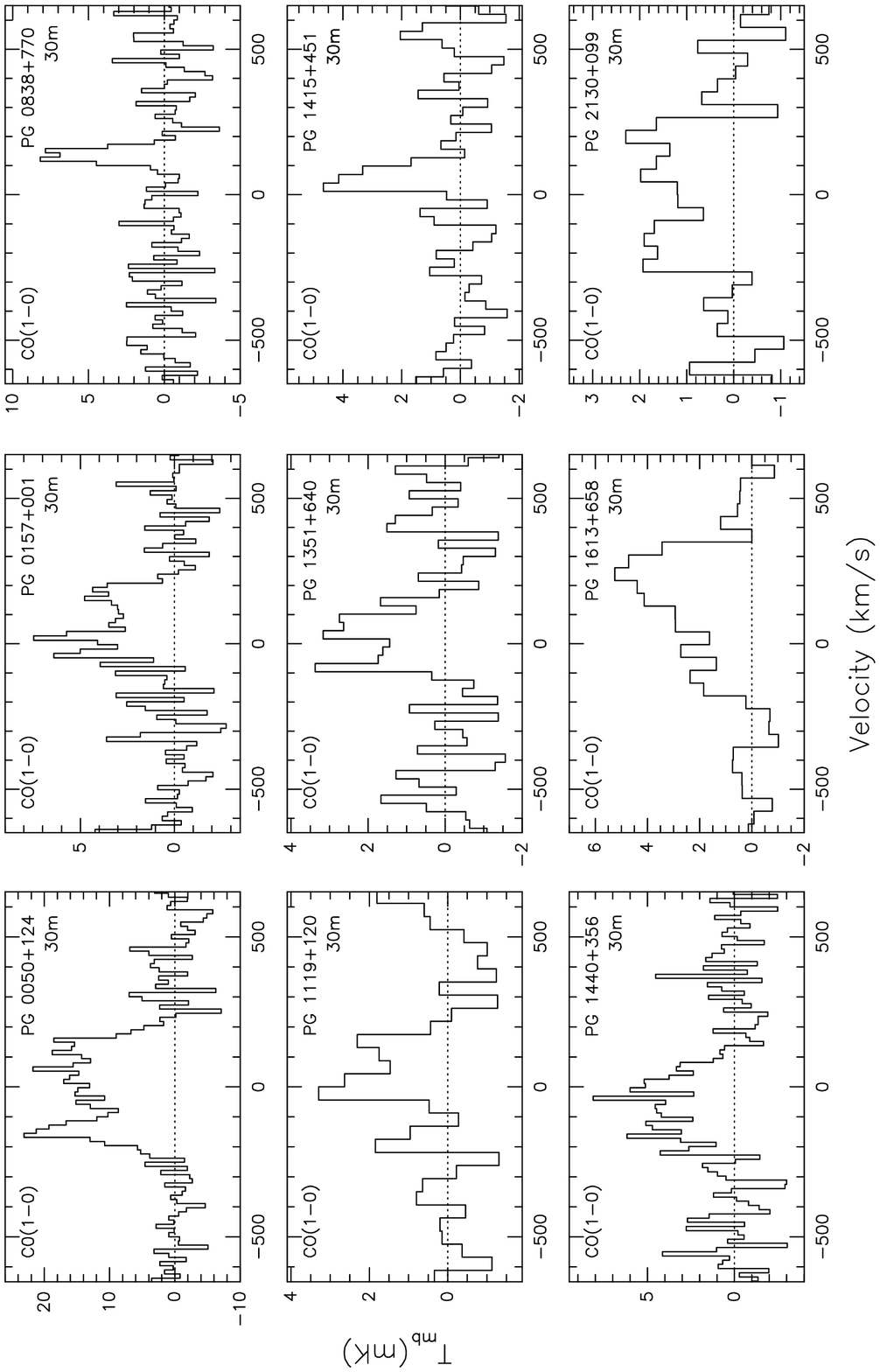}
\newpage
\plotone{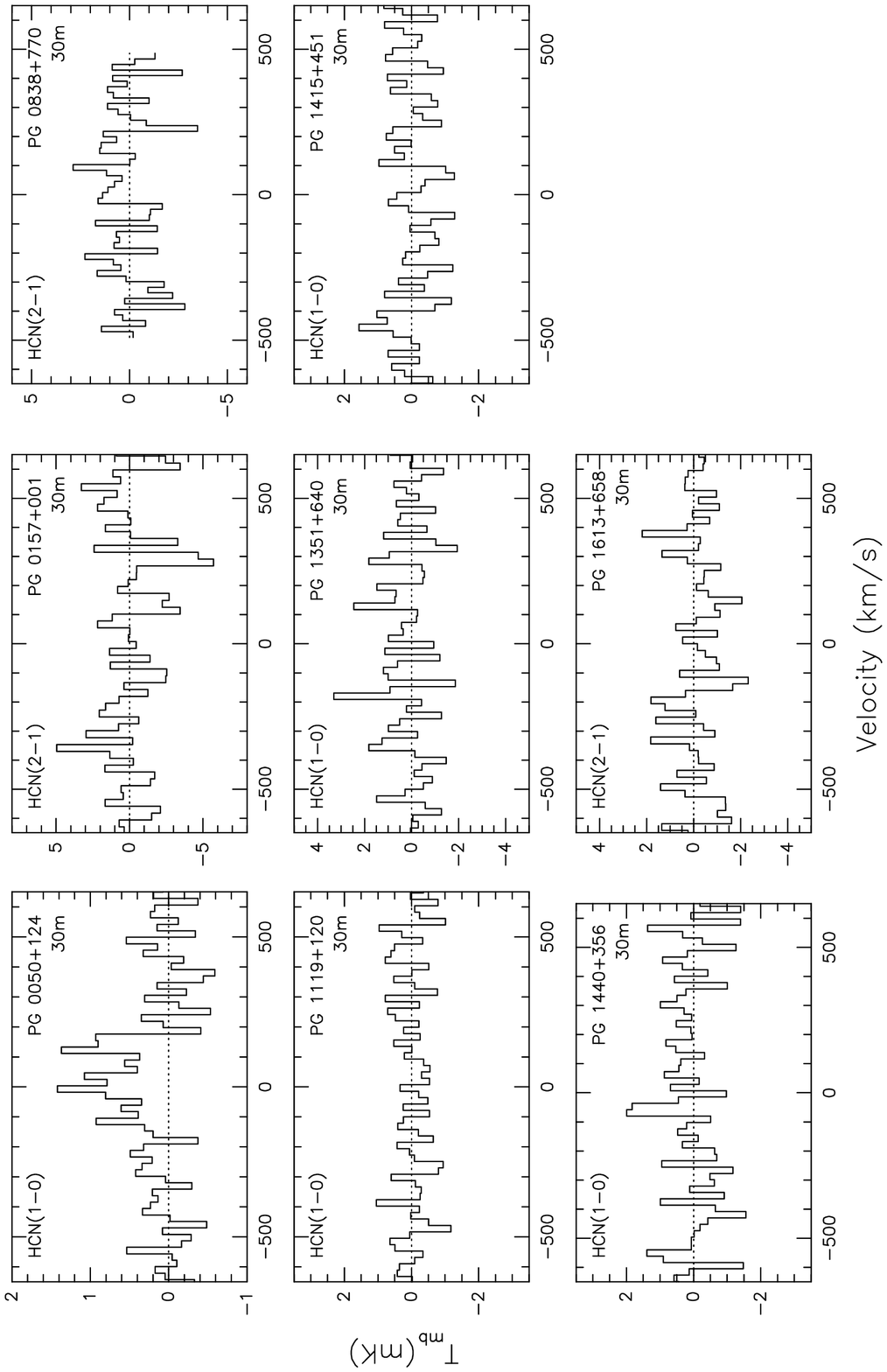}
\newpage
\plotone{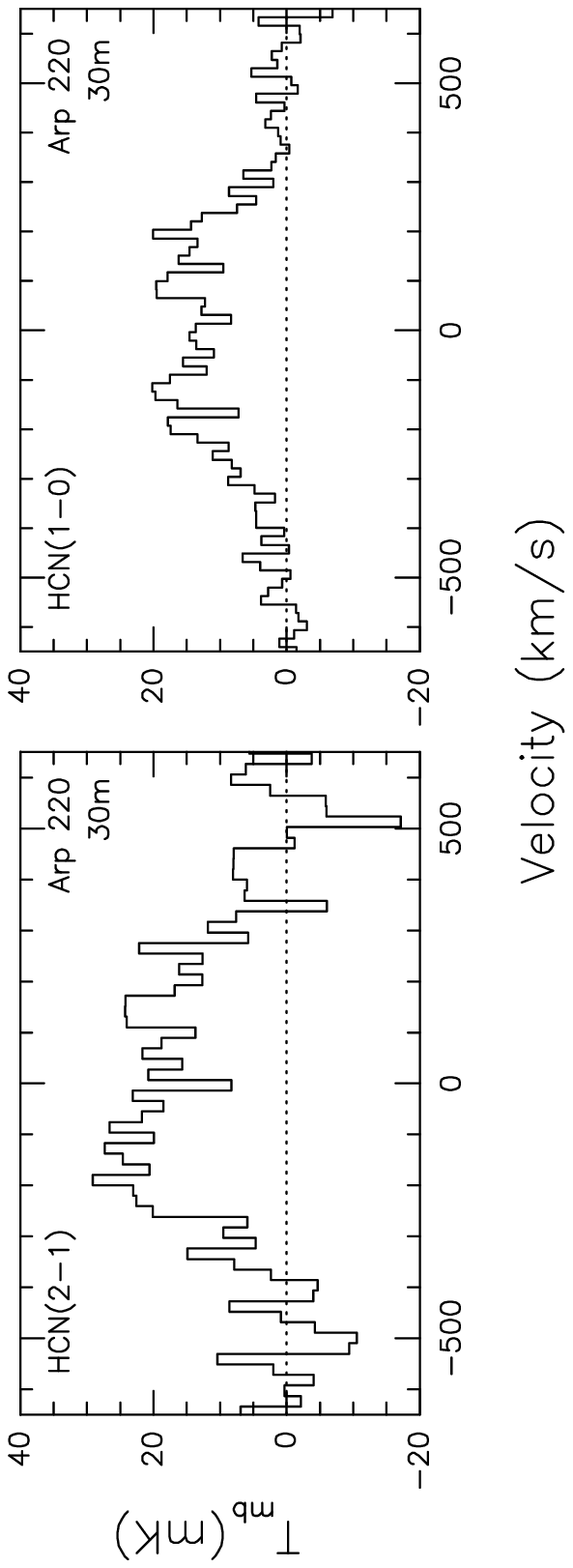}
\newpage
\epsscale{.8}
\plotone{figure4.ps}
\newpage
\plotone{figure5.ps}
\newpage
\plotone{figure6.ps}
\newpage
\plotone{figure7.ps}
\newpage
\plotone{figure8.ps}
\newpage
\plotone{figure9.ps}

\clearpage
%\documentstyle[apjpt4] {article}
%\begin{document}
%
\begin{deluxetable}{lcccrrrr}
\rotate
\tablenum{1}
\tablewidth{0pt}
\tablecaption{Source List}
\tablehead{
\multicolumn{1}{c}{} &
\multicolumn{2}{c}{Coordinates (J2000.0)} &
\multicolumn{1}{c}{$D_L^{~~~a}$} &
\multicolumn{1}{c}{} &
\multicolumn{1}{c}{$L_{\rm IR}^{~~~a}$} &
\multicolumn{1}{c}{$L_{\rm FIR}^{~~~a}$} &
\multicolumn{1}{c}{$SFR^{b}$}
\nl
\multicolumn{1}{c}{Source} &
\multicolumn{1}{c}{R.A.} &
\multicolumn{1}{c}{Decl.} &
\multicolumn{1}{c}{(Mpc)} &
\multicolumn{1}{c}{$z_{\rm CO}$} &
\multicolumn{1}{c}{($\times10^{11}L_\odot$)} &
\multicolumn{1}{c}{($\times10^{11}L_\odot$)} &
\multicolumn{1}{c}{M$_\odot$ yr$^{-1}$}
}
\startdata
PG 0050+124 = IZw1      & 00:53:34.94 & +12:41:36.20 & 250 & 0.061 & 7.2 & 2.8 & 95 \nl 
PG 0157+001 = Mrk 1014  & 01:59:50.21 & +00:23:40.62 & 680 & 0.163 & 32 & 21 & 420 \nl
PG 0838+770             & 08:44:45.36 & +76:53:09.20 & 540 & 0.131 & 2.9 & 1.4 & 40 \nl
PG 1119+120 = Mrk 734   & 11:21:47.15 & +11:44:19.00 & 200 & 0.050 & 1.1 & 0.44 & 15 \nl
PG 1351+640             & 13:53:15.83 & +63:45:45.60 & 360 & 0.088 & 5.6 & 2.1 & 75  \nl
PG 1415+451             & 14:17:00.84 & +44:56:06.50 & 470 & 0.114 & 1.6 & 0.66 & 20 \nl
PG 1440+356 = Mrk 478   & 14:42:07.48 & +35:26:23.10 & 320 & 0.078 & 2.8 & 1.4 & 40  \nl
PG 1613+658 = Mrk 876   & 16:13:57.21 & +65:43:10.60 & 530 & 0.129 & 7.6 & 3.9 & 100 \nl
PG 2130+099 = UGC 11763 & 21:32:27.77 & +10:08:19.50 & 250 & 0.063 & 2.1 & 0.62 & 30  \nl 
Average                                     &                        &                          &         &            &  &       & $90\pm125$ \nl
Median                                       &                        &                          &         &            &    &    &  40    \nl
\enddata
\tablenotetext{a}{Calculated using the prescription in Sanders \& Mirabel (1996), and 
assuming $H_0 = 75$ km s$^{-1}$ Mpc$^{-1}$
and $q_0 = 0.5$. The $L_{\rm IR}$ of PG 1119+120 and PG 2130+099 are
calculated using the flux densities published by Haas et al. (2003). For the remaining QSO
host galaxies, the 
flux densities published by Sanders et al. (1989) are used. }
\tablenotetext{b}{Calculated by assuming that $L_{\rm IR} =  L_{\rm starburst}$ (see \S 5).}
\end{deluxetable}
%
%\end{document}

%\documentstyle[apjpt4]{article}
%\begin{document}
\begin{deluxetable}{lllrr}
\pagestyle{empty}
\tablenum{2}
\tablewidth{0pt}
\tablecaption{Journal of Observations}
\tablehead{
\multicolumn{1}{c}{} &
\multicolumn{1}{c}{} &
\multicolumn{1}{c}{Date} &
\multicolumn{1}{c}{Time} &
\multicolumn{1}{c}{T$_{sys}$}  \nl
\multicolumn{1}{c}{Source} &
\multicolumn{1}{c}{Transition} &
\multicolumn{1}{c}{(mmm-yy)} &
\multicolumn{1}{c}{(hr)} &
\multicolumn{1}{c}{(K)}}
\startdata
PG 0050+124 & CO(1$\to0$) & Dec-03 & 40 & 161 \nl
            & HCN($1\to0$) & May-03 & 527 & 102  \nl
            & HCN($1\to0$) & Sep-03 & 215 & 111 \nl
PG 0157+001 & CO(1$\to0$) & Dec-03 & 50 & 111 \nl
            & HCN($1\to0$) & Dec-03 & 70 & 260  \nl
PG 0838+770 & CO(1$\to0$) & Dec-03 & 45 & 111 \nl
            & HCN($2\to1$) & May-03 & 220 & 217  \nl
PG 1119+120 & CO(1$\to0$) & Dec-03 & 67 & 95 \nl
            & HCN($1\to0$) & Jun-03 & 301 & 108 \nl
PG 1351+640 & CO(1$\to0$) & May-06 & 98 & 146 \nl
            & HCN($1\to0$) & Jul-03 & 300 & 111  \nl
PG 1415+451 & CO(1$\to0$) & Dec-03 & 50 & 106 \nl
            & HCN($1\to0$) & Dec-03 & 205 & 116 \nl
PG 1440+356 & CO(1$\to0$) & Dec-03 & 80 & 126 \nl
            & HCN($1\to0$) & Dec-03 & 210 & 120 \nl
PG 1613+658 & CO(1$\to0$) & Dec-03 & 35 & 126 \nl
            & HCN($2\to1$) & Dec-03 & 125 & 235  \nl
PG 2130+099 & CO(1$\to0$) & Dec-03 & 140 & 157 \nl

\enddata
\end{deluxetable}
%\end{document}

%\documentstyle[apjpt4] {article}
%\begin{document}
%
\begin{deluxetable}{lllrrrrrrr}
\rotate
\tablenum{3}
\tablewidth{0pt}
\tabletypesize{\scriptsize}
\tablecaption{Emission Line Properties}
\tablehead{
\multicolumn{1}{c}{Source} &
\multicolumn{1}{c}{$z_{\rm CO}$} &
\multicolumn{1}{c}{line} &
\multicolumn{1}{c}{$\Delta v_{\rm FWHM}$} &
\multicolumn{1}{c}{$T_{\rm mb} \Delta v$} &
\multicolumn{1}{c}{$S_{\rm line} \Delta v$} &
\multicolumn{1}{c}{${L'_{\rm line}}^{a,b}$} &
\multicolumn{1}{c}{$M$(H$_2)^{c}$} &
\multicolumn{1}{c}{${L_{\rm IR}^{~~d} \over L'_{\rm line}}$} &
\multicolumn{1}{c}{${L_{\rm IR}^{~~d,e} \over L'_{\rm HCN(1\to0)}}$} \nl
\multicolumn{1}{c}{} &
\multicolumn{1}{c}{} &
\multicolumn{1}{c}{} &
\multicolumn{1}{c}{(km s$^{-1}$)} &
\multicolumn{1}{c}{(K km s$^{-1}$)} &
\multicolumn{1}{c}{(Jy km s$^{-1}$)} &
\multicolumn{1}{c}{(K km s$^{-2}$ pc$^2$)} &
\multicolumn{1}{c}{($M_\odot$)} &
\multicolumn{1}{c}{} &
\multicolumn{1}{c}{} 
}
\startdata
\nl
PG 0050+124 & 0.061 & CO(1$\to$0) & 370 & 6.0$\pm0.2$ & 30$\pm1.1$ & $4.2\times10^9$ & $1.7\times10^{10}$ & 170 & \nodata  \nl
            & 0.061 & HCN(1$\to$0) & 290 & 0.27$\pm0.03$  & 1.3$\pm0.1$    & $3.2\times10^8$ & \nodata & 2260 & \nodata  \nl
PG 0157+001 & 0.163 & CO(1$\to$0) & 270 & 1.1$\pm0.1$ & 5.5$\pm0.5$ & $5.2\times10^9$ & $2.1\times10^{10}$ & 610 & \nodata  \nl 
            &       & HCN(1$\to$0) &   & $-0.12\pm0.16$  & $-0.60\pm0.80$    & $<9.6\times10^8$ & \nodata & $>$3300 & $>$1100  \nl
PG 0838+770 & 0.132 & CO(1$\to$0) & 60 & 0.50$\pm0.07$ & 2.5$\pm0.4$ & $1.5\times10^9$ & $6.1\times10^9$ & 190 &  \nl
            &       & HCN(2$\to$1) &    & $0.12\pm0.06$     & $0.60\pm0.31$     & $<1.8\times10^8$ & \nodata & $>$1590 & $>$530 \nl
PG 1119+120 & 0.050 & CO(1$\to$0) & 220 & 0.54$\pm0.09$ & 2.7$\pm0.4$ & $2.5\times10^8$ & $1.0\times10^9$ & 450 &   \nl
            &       & HCN(1$\to$0) &    & $-0.01\pm0.04$     & $-0.05\pm0.19$    & $<9.2\times10^7$ & \nodata & $>$1230 & \nodata \nl
PG 1351+640 & 0.088 & CO(1$\to$0) & 260 & 0.54$\pm$0.09  &  2.7$\pm0.5$    & $7.7\times10^8$ & $3.1\times10^9$ & 730 & \nodata   \nl
            &       & HCN(1$\to$0) &    & $0.09\pm0.08$     & $0.46\pm0.40$   & $<5.5\times10^8$ & \nodata & $>$1000 & \nodata \nl
PG 1415+451 & 0.114 & CO(1$\to$0) & 90 & 0.42$\pm0.06$ & 2.1$\pm0.3$ & $9.8\times10^8$ & $3.9\times10^9$ & 160 & \nodata \nl
            &       & HCN(1$\to$0) &    & $-0.02\pm0.04$    & $-0.12\pm0.20$    & $<3.8\times10^8$ & \nodata & $>$420& \nodata \nl
PG 1440+356 & 0.078 & CO(1$\to$0) & 310 & 1.3$\pm0.1$ & 6.6$\pm0.6$ & $1.5\times10^9$ & $2.0\times10^{9}$ & 190 & \nodata \nl
            &       & HCN(1$\to$0) &    & $0.09\pm0.08$    & $0.43\pm0.40$   & $<3.9\times10^8$ & \nodata & $>$720 & \nodata \nl
PG 1613+658 & 0.129 & CO(1$\to$0) & 400 & 1.6$\pm0.1$ & 8.0$\pm0.6$ & $4.8\times10^9$ & $1.9\times10^{10}$ & 160 & \nodata \nl
            &       & HCN(2$\to$1) &    & $-0.17\pm0.13$     & $-0.83\pm0.63$    & $<4.1\times10^8$ & \nodata & $>$1850 & $>$620 \nl
PG 2130+099 & 0.063 & CO(1$\to$0) & 530 & 0.78$\pm0.1$ & 3.9$\pm0.5$ & $5.6\times10^8$ & $2.2\times10^9$ & 380 & \nodata \nl
            &       &             &     &              &             &                 &                 &     &         \nl
 Arp 220 & 0.0181 & HCN(1$\to$0) &  550   &  $9.0\pm0.8$ & $45\pm3$ & $9.6\times10^8$ & \nodata & 1450 & \nodata \nl
                &               & HCN(2$\to$1) &   550   & $12\pm1.4$  & $60\pm7$  & $3.2\times10^8$ & \nodata & \nodata & \nodata \nl 
            &       &             &     &              &             &                 &                 &     &         \nl
PG QSO average (CO)&       &             & 260$\pm150$ &      &             &                 &                 \nl
\enddata
\tablenotetext{a}{$3\sigma$ upper limit.}
\tablenotetext{b}{Calculated assuming $H_0 = 75$ km s$^{-1}$ Mpc$^{-1}$
and $q_0 = 0.5$.}
\tablenotetext{c}{Calculated assuming $\alpha = 4 M_{\odot}$ (K km s$^{-1}$
pc$^2)^{-1}$.}
\tablenotetext{d}{In units of L$_\odot$ (K km s$^{-1}$ pc$^2$)$^{-1}$.} 
\tablenotetext{e}{ $L'_{\rm HCN(2\to1)} / L'_{\rm HCN(1\to0)}$ conversion is achieved by adopting
the $L'_{\rm HCN(1\to0)}/L'_{\rm HCN(2\to1)}$ (=2.997) of Arp 220. See \S 4.2.}
\end{deluxetable}
%
%\end{document}

%\documentstyle[apjpt4] {article}
%\begin{document}
%
\begin{deluxetable}{lllllllr}
\tablenum{4}
\tablewidth{0pt}
\tabletypesize{\scriptsize}
\tablecaption{Infrared-to-HCN Luminosity Ratio Comparisons and Adjusted Star Formation Rates}
\tablehead{
\multicolumn{1}{c}{Source} &
\multicolumn{1}{c}{Type} &
\multicolumn{1}{c}{Class$^{\rm a}$} &
\multicolumn{1}{c}{${L_{\rm IR} \over L'_{\rm HCN}}$$^{\rm b}$} &
\multicolumn{1}{c}{${L_{\rm FIR} \over L'_{\rm HCN}}$$^{\rm b}$} &
\multicolumn{1}{c}{$\left<L_{\rm IR} / L'_{\rm HCN}\right>_{\rm cool} \over L_{\rm IR} / L'_{\rm HCN}$} &
\multicolumn{1}{c}{$\left<L_{\rm FIR} / L'_{\rm HCN}\right>_{\rm cool} \over L_{\rm FIR} / L'_{\rm HCN}$} &
\multicolumn{1}{c}{SFR$^{\rm d}$} 
}
\startdata
median  &  cool  & \nodata & $890^{+440}_{-470}$ & $660^{+370}_{-330}$ & \nodata & \nodata & \nodata \nl
median & warm  &  \nodata & $1850^{+1240}_{-1070}$ & $1220^{+860}_{-810}$ & $0.48\pm0.25$ & $0.54\pm0.28$ & \nodata \nl
median & QSO & \nodata     & $3150^{+6300}_{-1050}$ & $1575^{+4720}_{-530}$ & $0.28^{+0.35}_{-0.24}$ & $0.42^{+0.56}_{-0.35}$ & \nodata \nl
\hline
PG 1351+640 & QSO & Seyfert 1 & 12120$^{+24230}_{-4040}$ & 4540$^{+9090}_{-1520}$ & $0.07^{0.09}_{-0.06}$$^{\rm c}$ & $0.15^{+0.19}_{-0.12}$$^{\rm c}$ & $5^{+7}_{-5}$\nl
PG 0157+001 & QSO & Seyfert 1 & 10120$^{+20230}_{-3370}$ & 6640$^{+13280}_{-2210}$ & $0.09^{+0.11}_{-0.07}$$^{\rm c}$ & $0.10^{+0.13}_{-0.08}$$^{\rm c}$ & $37^{+46}_{-31}$ \nl
PG 1119+120 & QSO & Seyfert 1 & 7530$^{+15070}_{-2510}$ &3010$^{+6030}_{-1000}$ & $0.12^{+0.15}_{-0.10}$$^{\rm c}$ & $0.22^{+0.29}_{-0.18}$$^{\rm c}$ & $2^{+2}_{-1}$ \nl
Cloverleaf Quasar & high-$z$ AGN & Seyfert 1& 7700 & 1700 & $0.12\pm0.06$ & $0.39\pm0.20$ & $375\pm190$ \nl
PG 2130+099 & QSO & Seyfert 1 & 6250$^{+12500}_{-2080}$ & 1840$^{+3690}_{-610}$ & $0.14^{+0.18}_{-0.12}$$^{\rm c}$ & $0.36^{+0.48}_{-0.30}$$^{\rm c}$ &  $4^{+5}_{-3}$ \nl
IRAS FSC10214+4724 & high-$z$ AGN & Seyfert 2 & 4800 & 2600 & $0.19\pm0.10$ & $0.25\pm0.13$ & $223\pm110$ \nl
NGC 4418 & warm & LINER& 4610 & 3830 & $0.19\pm0.10$ & $0.17\pm0.09$ & $3\pm2$ \nl
PG 1440+356 &QSO & Seyfert 1 & 3150$^{+6300}_{-1050}$ & 1580$^{+3150}_{-520}$ & $0.28^{+0.35}_{-0.24}$$^{\rm c}$ & $0.42^{+0.56}_{-0.35}$$^{\rm c}$ & $10^{+13}_{-9}$ \nl
PG 0838+770 & QSO & Seyfert 1 & 3150$^{+6300}_{-1050}$ & 1520$^{+3040}_{-510}$ & $0.28^{+0.35}_{-0.24}$$^{\rm c}$ & $0.43^{+0.58}_{-0.36}$$^{\rm c}$ & $11^{+13}_{-9}$ \nl
PG 1415+451 & QSO & Seyfert 1 & 2720$^{+5430}_{-910}$ & 1120$^{+2240}_{-370}$ & $0.33^{+0.41}_{-0.28}$$^{\rm c}$ & $0.59^{+0.79}_{-0.49}$$^{\rm c}$ & $7^{+9}_{-6}$ \nl
PG 1613+658 & QSO & Seyfert 1 & 2630$^{+5270}_{-880}$ & 1380$^{2760}_{-460}$ & $0.34^{+0.42}_{-0.28}$$^{\rm c}$ & $0.48^{+0.64}_{-0.40}$$^{\rm c}$ & $34^{+42}_{-29}$ \nl
NGC 1614 & warm & HII/LINER & 3090 & 2090 & $0.29\pm0.15$ & $0.32\pm0.17$ & $15\pm7$ \nl
PG 0050+124 = IZw1 & QSO & Seyfert 1 & 2260 & 880 & $0.39\pm0.20$ & $0.75\pm0.39$ & $37\pm19$  \nl 
IRAS FSC05189-2524 & Seyfert 2 & warm & 1900 & 1260 & $0.47\pm0.24$ & $0.52\pm0.28$ & $73\pm37$ \nl
NGC 7469 & warm & Seyfert 1 & 1860 & 1190 & $0.48\pm0.25$ & $0.55\pm0.29$ & $26\pm13$ \nl
NGC 3034 = M82 & warm & HII & 1850 & 1270 & $0.48\pm0.25$ & $0.52\pm0.27$ & $3\pm2$ \nl
Mrk 231 & warm & Seyfert 1 & 1630 & 1030 & $0.55\pm0.28$ & $0.64\pm0.34$ & $219\pm112$ \nl
NGC 1068 & warm & Seyfert 2 & 780 & 320 & $1.1\pm0.59$ & $2.0\pm1.1$ & $42\pm21$ \nl
NGC 7479 & warm & Seyfert 2 & 660 & 410 & $1.4\pm0.70$ & $1.6\pm0.83$ & $13\pm7$ \nl
\enddata
\tablenotetext{a}{References for optical spectral classification: Veilleux et al. (1995) and Ho, Filippenko,
\& Sargent (1997). }
\tablenotetext{b}{In units of L$_\odot$ (K km s$^{-1}$ pc$^2$)$^{-1}$.}
\tablenotetext{c}{Calculated by adopting $L'_{\rm HCN}  / L'_{\rm CO} = 0.06$ (see \S 5.1).}
\tablenotetext{d}{Star formation rate in M$_\odot$ yr$^{-1}$.}
\end{deluxetable}
%
%\end{document}

\end{document}